\documentclass[a4paper,20pt,oneside,Palatino]{article}
	\usepackage{latexsym}
	\usepackage[dvips]{graphics}
	\usepackage[dvips]{color}
	\usepackage{graphicx}
	\usepackage{simplemargins}	

	\setleftmargin{1in} 
	\setrightmargin{1in} 
	\settopmargin{1in}
	\setbottommargin{0.8in}

\begin{document}
	\begin{center}
	{\large\bf Circumventing Astrophysical Bounds Upon PVLAS Experiment}
		
	\bigskip
	Subhayan Mandal and Pankaj Jain\\
	
	Physics Department\\
	I.I.T. Kanpur\\
	Kanpur, India 208016\\
	email:shabsslg@iitk.ac.in, pkjain@iitk.ac.in
	\end{center}
	
	\bigskip

	\begin{minipage}[C]{430pt}
	\noindent
	{\bf Abstract:} We study the possibility of evading astrophysical bound on light pseudoscalars. We argue that the solar bounds can be evaded if we have a sufficiently strong self-coupling of the pseudoscalars. The required coupling does not conflict with any known experimental bounds. We also show that it is possible to find a coupling range such that the results of the recent PVLAS experiment are not in conflict with any astrophysical bounds.
	\end{minipage}

	\section{Introduction}

	\noindent
	Recently the PVLAS collaboration \cite{zav} studied the polarization of light propagating through transverse magnetic field. They found rotation of polarization as well as induction of ellipticity in the emergent light. To expound this otherwise inexplicable phenomena we hypothesize the existence of light pseudoscalar particles. The PVLAS experimental \cite{zav} result can be explained by the mixing of photons with pseudoscalars mediated by magnetic field. From the experimental parameters we can easily derive the required mass of pseudoscalar particles, and  it's coupling to photon, $g_{a\gamma\gamma}$. The derived mass and coupling are well above the standard astrophysical limits \cite{astrobounds} put on the standard $\phi$ particle called `QCD Axion' and other hypothesized `Pseudoscalars'. The astrophysical limits are derived by requiring that pseudoscalar do not cool the stellar body too quickly. Assuming the coupling range predicted by the PVLAS experiment, it is found that objects such as sun would have a very small lifetime due to rapid loss of energy due to pseudoscalar emission. We note here that the derived coupling and mass do not violet any laboratory experiment \cite{cameron}. Some work has been done to see if the astrophysical limits can somehow be evaded [4-13]. Proposals for new experiments \cite{Jaeckel, Koetz} and observation \cite{Fairbairn} have also surfaced. Furthermore new constraints have been imposed on axion monopole-dipole coupling \cite{Baessler}. In this brief resume we shall closely follow our recent paper \cite{mandal}, in which it is shown that both the astrophysical bounds \cite{astrobounds} and the CAST experiment \cite{CAST} can be consistent with the PVLAS data \cite{zav}.

 	\section{Self Coupling Of Pseudoscalars}

	\noindent
	We seek solution to the above problem by assuming large self coupling of the $\phi$ particles. We propose  to restrict the 'Mean Free Path' of the $\phi$ particles by introducing a strong interaction within themselves. This will in turn trap the $\phi$ particles within the stellar plasma despite their interaction to visible matter being very weak. Hence their is no outflow of energy via $\phi$ channel. This allows us to evade astrophysical bounds on pseudoscalars.

	\medskip
	
	\noindent
	We first argue that in the primitive stage of the sun there will be production of pseudoscalars inside the solar core due to compton or primakoff type of processes. Then we assume that the self coupling constant is large but still lies in the perturbative regime. In a conservative estimate of the pseudoscalar density inside the sun we assume that the total $\phi$ luminosity of the sun is about half of the $\gamma$'s. Using that value we would be able to calculate the number density of the $\phi$'s, assuming steady state.
	
	\medskip

	\noindent
	Using these two assumptions we calculate the mean free path due to self interaction of $\phi$ particles. If that is found to be much less than the solar radius then they will start accumulating inside the sun. 
	
	\section{Hypothesis Testing}

	\noindent
	The interaction Lagrangian of pseudoscalars,

	\begin{equation}
	\mathcal{L} = {\phi\over4M}F_{\mu\nu}\tilde{F}^{\mu\nu} + {1\over4!}\lambda\phi^4
	\end{equation} 
	
	\noindent
	The cross section of the two by two process $\phi(k_1)\phi(k_2) \rightarrow \phi(k_1^{'})\phi(k_2^{'})$ due to the last term in the above interaction Lagrangian is

	\begin{equation}
	\sigma_{\phi\phi} = \frac{\lambda^2}{32\pi E_{cm}^2}
	\end{equation}

	\noindent
	Taking $\lambda$ to be $1$ \& $E_{cm}$ = $1~{\rm KeV}$ we find the mean free path,
	\begin{equation}
	l_{\phi\phi} = {1\over n_{\phi}\sigma{\phi\phi}} = 3\times10^7~{\rm cm}
	\end{equation}

	\noindent
	This is roughly three orders of magnitude smaller than the solar radius, $R_{\odot}$. So pseudoscalars will accumulate inside the sun, eventually attaining a steady state with photons

	\section{Fragmentation}
	
	\noindent
	Another process that will occur inside solar plasma is shown in fig.~1, namely, $\phi(k_1)\phi(k_2) \rightarrow \phi(k_1^{'})\phi(k_2^{'})\phi(k_3^{'})\phi(k_4^{'})$. This process is suppressed from the two by two process by only two powers of coupling and can contribute significantly.

	\begin{center}
 	\scalebox{0.7}{\begin{picture}(0,0)%
\includegraphics{frag.pstex}%
\end{picture}%
\setlength{\unitlength}{3947sp}%
\begingroup\makeatletter\ifx\SetFigFont\undefined%
\gdef\SetFigFont#1#2#3#4#5{%
  \reset@font\fontsize{#1}{#2pt}%
  \fontfamily{#3}\fontseries{#4}\fontshape{#5}%
  \selectfont}%
\fi\endgroup%
\begin{picture}(4950,2179)(226,-1853)
\put(2101,-1786){\makebox(0,0)[lb]{\smash{{\SetFigFont{14}{16.8}{\sfdefault}{\mddefault}{\updefault}{\color[rgb]{0,0,1}$\phi(k_1^{'})$}%
}}}}
\put(5176,-736){\makebox(0,0)[lb]{\smash{{\SetFigFont{14}{16.8}{\sfdefault}{\mddefault}{\updefault}{\color[rgb]{0,0,1}$\phi(k_3^{'})$}%
}}}}
\put(4426,-1711){\makebox(0,0)[lb]{\smash{{\SetFigFont{14}{16.8}{\sfdefault}{\mddefault}{\updefault}{\color[rgb]{0,0,1}$\phi(k_2^{'})$}%
}}}}
\put(226,-1711){\makebox(0,0)[lb]{\smash{{\SetFigFont{14}{16.8}{\sfdefault}{\mddefault}{\updefault}{\color[rgb]{1,0,0}$\phi(k_2)$}%
}}}}
\put(4501,-61){\makebox(0,0)[lb]{\smash{{\SetFigFont{14}{16.8}{\sfdefault}{\mddefault}{\updefault}{\color[rgb]{0,0,1}$\phi(k_4^{'})$}%
}}}}
\put(976,-61){\makebox(0,0)[lb]{\smash{{\SetFigFont{14}{16.8}{\sfdefault}{\mddefault}{\updefault}{\color[rgb]{1,0,0}$\phi(k_1)$}%
}}}}
\end{picture}%
}
	\end{center}

	\centerline{Fig. 1: Fragmentation Process,  $\phi(K_1)\phi(K_2)\rightarrow\phi(K_1^{'})\phi(K_2^{'})\phi(K_3^{'})\phi(K_4^{'})$}

	\medskip

	\noindent
	This kind of fragmentation reaction of $\phi$ particles would degrade their energy per particle until it's energy is of the order of it's rest mass. Since the initial energy of the $\phi$ particles were around 1 KeV, commensurate with the solar core temperature, and their rest mass is only 1 meV [from PVLAS], this fragmentation would lead us to the enhancement in the number density of $\phi$'s by a factor of $10^6$. This will reduce the mean free path of the two by two process $\phi(k_1)\phi(k_2) \rightarrow \phi(k_1^{'})\phi(k_2^{'})$ to \textsf{ 10 cm}. But the above is an underestimate only due to the $E_{cm}$ dependence of $\sigma_{\phi\phi}$ The mean free path of the above reaction would then further reduce to \textsf{10$^{-5}$ cm.}

	\section{Steady State}

	\noindent
	Pseudoscalar particles eventually reach a steady state with photons. The diagrams for this are shown below, in fig.~2.

	\begin{center}
 	\scalebox{0.5}{\begin{picture}(0,0)%
\includegraphics{Prim.pstex}%
\end{picture}%
\setlength{\unitlength}{3947sp}%
\begingroup\makeatletter\ifx\SetFigFont\undefined%
\gdef\SetFigFont#1#2#3#4#5{%
  \reset@font\fontsize{#1}{#2pt}%
  \fontfamily{#3}\fontseries{#4}\fontshape{#5}%
  \selectfont}%
\fi\endgroup%
\begin{picture}(10033,3406)(289,-2837)
\put(301,-2086){\makebox(0,0)[lb]{\smash{{\SetFigFont{14}{16.8}{\sfdefault}{\mddefault}{\updefault}{\color[rgb]{0,0,1}$e^{-}(p)$}%
}}}}
\put(3976,-2086){\makebox(0,0)[lb]{\smash{{\SetFigFont{14}{16.8}{\sfdefault}{\mddefault}{\updefault}{\color[rgb]{0,0,1}$e^{-}(p{'})$}%
}}}}
\put(1351,-811){\makebox(0,0)[lb]{\smash{{\SetFigFont{14}{16.8}{\sfdefault}{\mddefault}{\updefault}{\color[rgb]{1,0,0}D(Q)}%
}}}}
\put(301,389){\makebox(0,0)[lb]{\smash{{\SetFigFont{14}{16.8}{\sfdefault}{\mddefault}{\updefault}{\color[rgb]{1,0,0}$\gamma(k)$}%
}}}}
\put(3976, 89){\makebox(0,0)[lb]{\smash{{\SetFigFont{14}{16.8}{\sfdefault}{\mddefault}{\updefault}{\color[rgb]{1,0,0}$\phi(k^{'})$}%
}}}}
\put(9076,314){\makebox(0,0)[lb]{\smash{{\SetFigFont{14}{16.8}{\sfdefault}{\mddefault}{\updefault}{\color[rgb]{1,0,0}$\gamma(k)$}%
}}}}
\put(5851, 89){\makebox(0,0)[lb]{\smash{{\SetFigFont{14}{16.8}{\sfdefault}{\mddefault}{\updefault}{\color[rgb]{1,0,0}$\phi(k^{'})$}%
}}}}
\put(6076,-2161){\makebox(0,0)[lb]{\smash{{\SetFigFont{14}{16.8}{\sfdefault}{\mddefault}{\updefault}{\color[rgb]{0,0,1}$e^{-}(p)$}%
}}}}
\put(9076,-2086){\makebox(0,0)[lb]{\smash{{\SetFigFont{14}{16.8}{\sfdefault}{\mddefault}{\updefault}{\color[rgb]{0,0,1}$e^{-}(p{'})$}%
}}}}
\put(7051,-811){\makebox(0,0)[lb]{\smash{{\SetFigFont{14}{16.8}{\sfdefault}{\mddefault}{\updefault}{\color[rgb]{1,0,0}D(Q)}%
}}}}
\put(3376,-2761){\makebox(0,0)[lb]{\smash{{\SetFigFont{14}{16.8}{\sfdefault}{\mddefault}{\updefault}{\color[rgb]{0,0,0}Direct \& Inverse Primakoff Processes}%
}}}}
\end{picture}%
}
	\end{center}

	\noindent
	The $\phi$ produced by the primakoff process will have roughly $10^6$ times the energy of the $\gamma$ produced by the inverse process. The production rate of $\phi$'s from $\gamma$'s per unit volume is $\sigma_{\gamma X \rightarrow \phi X} n_{\gamma}n_{X} v$ where v = c is the 'Speed Of Light'. We set this production rate to be $10^{-6}$ times the production rate of $\gamma$'s from $\phi$'s, namely, $\sigma_{\phi X \rightarrow \gamma X} n_{\phi}n_{X} v$. Since the direct process and the inverse process have almost the same cross section hence the number density of $\phi$ particles would be $10^6$ times as much as the $\gamma$ particles, which is of the order of $10^{23}$ per cm$^3$. This makes the mean free path of the $\phi(k_1)\phi(k_2) \rightarrow \phi(k^{'}_1)\phi(k^{'}_2)$ process be \textsf{10$^{-17}$ cm}.

	\medskip	

	\noindent
	Hence the contribution of $\phi$ particles to the radiative transport inside the Sun would be negligible compared to the contribution of $\gamma$'s, once Sun becomes the main sequence star, i.e. if we can employ the steady state condition.

	\section{Distribution Of $\phi$ Particles}

	\noindent
	So far we have only considered the $\phi$ distribution inside the core of the Sun, assuming steady state conditions. We also expect the $\phi$ density to be significant outside the core, extending much beyond the solar radius. This pseudoscalar halo will be gravitationally bound to the Sun. We assume this halo to be spherically symmetric since we expect it's rotational speed to be small. We also expect that the velocity of the $\phi$ particles inside the core of the 'Sun' to be much greater than the escape velocity from the surface of the sun $v_{\odot} \approx 6\times 10^5m/s$. But as they propagate towards the edge of the Sun they eventually become nonrelativistic, due the the energy loss processes, shown in Fig. 3.

	\begin{center}
	\scalebox{0.45}{\begin{picture}(0,0)%
\includegraphics{elc.pstex}%
\end{picture}%
\setlength{\unitlength}{3947sp}%
\begingroup\makeatletter\ifx\SetFigFont\undefined%
\gdef\SetFigFont#1#2#3#4#5{%
  \reset@font\fontsize{#1}{#2pt}%
  \fontfamily{#3}\fontseries{#4}\fontshape{#5}%
  \selectfont}%
\fi\endgroup%
\begin{picture}(6109,2472)(118,-1717)
\put(151,-1636){\makebox(0,0)[lb]{\smash{{\SetFigFont{20}{24.0}{\sfdefault}{\mddefault}{\updefault}{\color[rgb]{1,0,0}Energy Loss Due To  Compton Coupling Term }%
}}}}
\put(1201,539){\makebox(0,0)[lb]{\smash{{\SetFigFont{20}{24.0}{\sfdefault}{\mddefault}{\updefault}{\color[rgb]{1,0,0}$\mathcal{L}_I = g_A\bar{\psi}\gamma_5\psi\phi$ }%
}}}}
\end{picture}%
} \scalebox{0.35}{\begin{picture}(0,0)%
\includegraphics{elpv.pstex}%
\end{picture}%
\setlength{\unitlength}{3947sp}%
\begingroup\makeatletter\ifx\SetFigFont\undefined%
\gdef\SetFigFont#1#2#3#4#5{%
  \reset@font\fontsize{#1}{#2pt}%
  \fontfamily{#3}\fontseries{#4}\fontshape{#5}%
  \selectfont}%
\fi\endgroup%
\begin{picture}(5766,2424)(418,-2119)
\put(826, 89){\makebox(0,0)[lb]{\smash{{\SetFigFont{20}{24.0}{\sfdefault}{\mddefault}{\updefault}{\color[rgb]{1,0,0}$L_I = {1\over4!}\lambda\phi^4$}%
}}}}
\end{picture}%
} 
	\end{center}
	
	\centerline{Fig. 3: Energy Loss Processes}

	\medskip

	\noindent
	Using the presuppositions stated in section 3 and employing the steady state conditions we may estimate the density profile of the $\phi$ halo for r $>$ R$_{\odot}$. The pressure of $\phi$'s at a distance $r$ from the center of the Sun is $P = \rho_{\phi}<v^2>/3$. And the mean velocity at a distance $r$ is given by

	\begin{equation}
	<v^2> = C + \frac{2G(M_{\odot} + M_r)}{r} 
	\end{equation}

	\noindent
	where $C$ = $<v_0^2>$ - $\frac{2G(M_{\odot} + M_{R_{\odot}})}{R_{\odot}}$ is a constant. M$_{\odot}$ is the mass of visible matter of the Sun \& $M_r$ is the total mass of the $\phi$ particles within a sphere of radius $r$ centered at the center of the Sun.

	\medskip

	\noindent
	We also have 

	\begin{equation}
	\frac{dM_r}{dr} = 4\pi r^2 \rho_{\phi}(r)
	\end{equation}
	
	\noindent
	and,

	\begin{equation}
	\frac{dP}{dr} = - {G(M_{\odot} + M_r)\rho_{\phi}\over r^2}
	\end{equation}

	\noindent
	Using equation 4, 5, 6 and the equation $P = \rho_{\phi}<v^2>/3$, we find,

	\begin{equation}
	\frac{d\rho_{\phi}}{dr} = - {1\over<v^2>}\left[{G(M_{\odot} + M_r)\rho_{\phi}\over r^2} + 8\pi G \rho_{\phi}^2 r \right ]
	\end{equation}

	\noindent
	Since the right hand side of this equation is negative for all $r$, we see $\rho_{\phi}$ decreases with $r$. In the small $r$ regime, where $M_r << M_{\odot}$ \& $8\pi G \rho_{\phi}^2 r^3 << M_{\odot} $ we find the asymptotic solution $\rho_{\phi}(r)/\rho_{\phi}(r_0)$ $ \approx $ $\sqrt{v^2(r)/v^2(r_0)}$. In the opposite asymptotic limit of large r, where $M_r >> M_{\odot}$ we obtain $\rho{\phi} \propto 1/r^2$ \& hence $M_r \propto r$.

	\medskip
	
	\noindent
	We numerically integrate the two [Equation 5 \& 6] first order linear coupled differential equation. The results are shown in Fig. 4.

	\begin{center}
	\includegraphics[scale = 0.5]{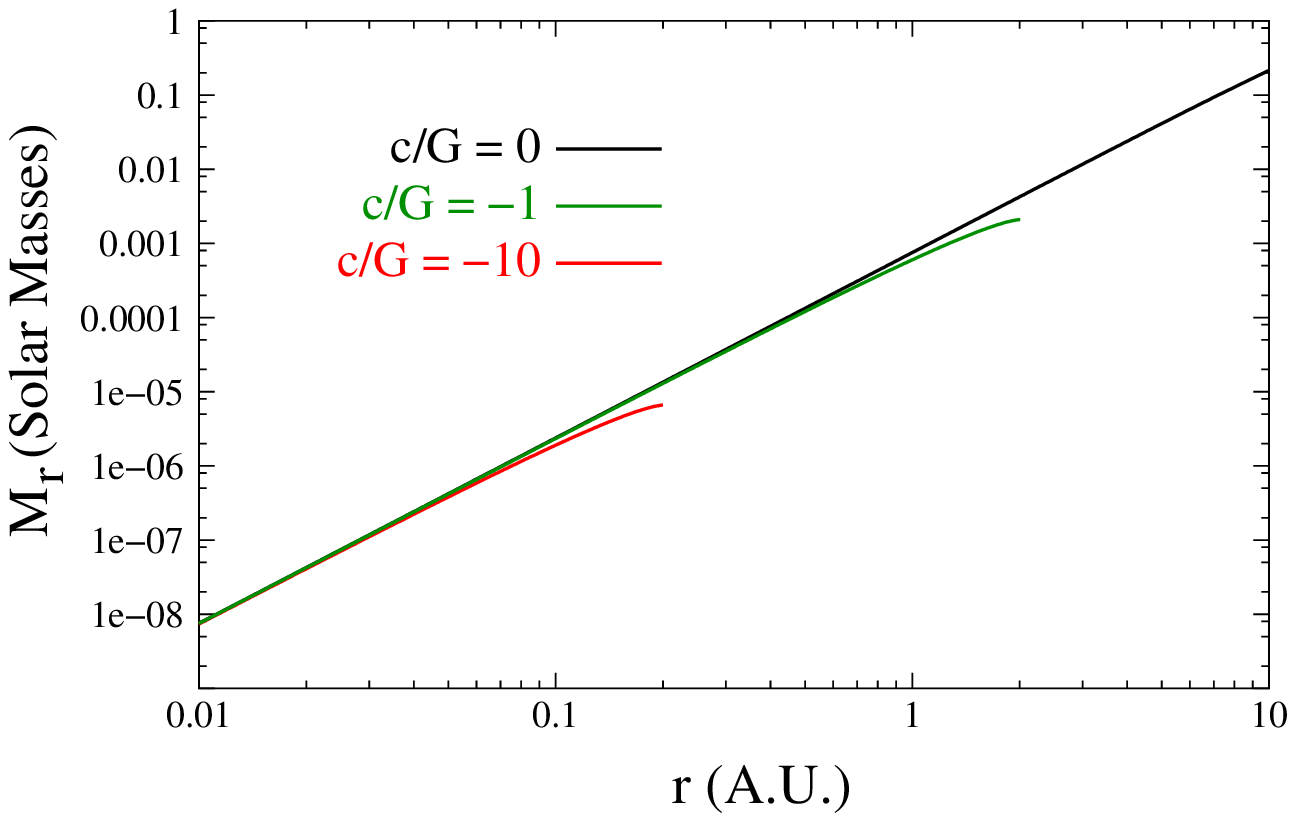}
	\end{center}

	\centerline{Fig. 4: Variation Of Pseudoscalar Mass With Distance}
	
	\medskip

	\noindent
	The results are shown for the different values of the parameter $c/G$. For $c/G \ge 0$ the $\phi$'s will have enough energy to escape the gravitational attraction of the sun \& hence the halo would extend to infinite distance. For values $c/G \le -1$ we find that the radius of the $\phi$ halo is quite small and contributes negligibly to the mass of the solar system. Hence, we see a wide range of parameters to evade all the astrophysical bounds on light pseudoscalars.

	\end{document}